\documentclass[journal=jpccck,manuscript=article]{achemso}

\usepackage[version=3]{mhchem} 

\renewcommand{\titlesize}{\Large}
\author{Ilana J. Porter}
\affiliation[CalChem]{Department of Chemistry, University of California, Berkeley, CA 94720, USA.}
\alsoaffiliation[lbnl]{Chemical Sciences Division, Lawrence Berkeley National Laboratory, Berkeley, CA 94720, USA.}
\altaffiliation{The authors contributed equally to this work.}
\author{Angela Lee}
\affiliation[CalChem]{Department of Chemistry, University of California, Berkeley, CA 94720, USA.}
\altaffiliation{The authors contributed equally to this work. Current address: Department of Chemistry, Massachusetts Institute of Technology, Cambridge, MA 02139, USA.}
\author{Scott K. Cushing}
\affiliation[CalChem]{Department of Chemistry, University of California, Berkeley, CA 94720, USA.}
\alsoaffiliation[lbnl]{Chemical Sciences Division, Lawrence Berkeley National Laboratory, Berkeley, CA 94720, USA.}
\altaffiliation{Current address: Division of Chemistry and Chemical Engineering, California Institute of Technology, Pasadena, CA, 91125, USA.}
\author{Hung-Tzu Chang}
\affiliation[CalChem]{Department of Chemistry, University of California, Berkeley, CA 94720, USA.}
\author{Justin C. Ondry}
\affiliation[CalChem]{Department of Chemistry, University of California, Berkeley, CA 94720, USA.}
\alsoaffiliation[Kavli]{Kavli Energy NanoScience Institute, Berkeley, CA 94720, USA.}
\altaffiliation{Current address: Department of Chemistry, James Franck Institute, and Pritzker School of Molecular Engineering, University of Chicago, Chicago, Illinois 60637, USA.}
\author{A. Paul Alivisatos}
\affiliation[CalChem]{Department of Chemistry, University of California, Berkeley, CA 94720, USA.}
\alsoaffiliation[Kavli]{Kavli Energy NanoScience Institute, Berkeley, CA 94720, USA.}
\alsoaffiliation[matlbnl]{Materials Sciences Division, Lawrence Berkeley National Laboratory, Berkeley, CA 94720, USA.}
\alsoaffiliation[calmse]{Department of Materials Science and Engineering, University of California, Berkeley, CA 94720, USA}
\author{Stephen R. Leone}
\email{srl@berkeley.edu}
\affiliation[CalChem]{Department of Chemistry, University of California, Berkeley, CA 94720, USA.}
\alsoaffiliation[lbnl]{Chemical Sciences Division, Lawrence Berkeley National Laboratory, Berkeley, CA 94720, USA.}
\alsoaffiliation[calphys]{Department of Physics, University of California, Berkeley, CA 94720, USA.}

\title{Characterization of Carrier Cooling Bottleneck in Silicon Nanoparticles by Extreme Ultraviolet (XUV) Transient Absorption Spectroscopy}
\titlesize{\huge}

\begin{document}

%
%
%
%
%

\begin{abstract}
  Silicon nanoparticles have the promise to surpass the theoretical efficiency limit of single-junction silicon photovoltaics by the creation of a ``phonon bottleneck", a theorized slowing of the cooling rate of hot optical phonons that in turn reduces the cooling rate of hot carriers in the material. To verify the presence of a phonon bottleneck in silicon nanoparticles requires simultaneous resolution of electronic and structural changes at short timescales. Here, extreme ultraviolet transient absorption spectroscopy is used to observe the excited state electronic and lattice dynamics in polycrystalline silicon nanoparticles following 800 nm photoexcitation, which excites carriers with 0.35 $\pm$ 0.03 eV excess energy above the $\Delta_1$ conduction band minimum. The nanoparticles have nominal 100 nm diameters with crystalline grain sizes of about $\sim$16 nm. The extracted carrier-phonon and phonon-phonon relaxation times of the nanoparticles are compared to those for a silicon (100) single crystal thin film at similar carrier densities ($2 \times 10^{19}$ cm$^{-3}$ for the nanoparticles and $6 \times 10^{19}$ cm$^{-3}$ for the film). The measured carrier-phonon and phonon-phonon scattering lifetimes for the polycrystalline nanoparticles are 870 $\pm$ 40 fs and 17.5 $\pm$ 0.3 ps, respectively, versus 195 $\pm$ 20 fs and 8.1 $\pm$ 0.2 ps, respectively, for the silicon thin film. The reduced scattering rates observed in the nanoparticles are consistent with the phonon bottleneck hypothesis.
\end{abstract}

\section{Introduction}
Silicon is a ubiquitous material in the solar energy industry, providing an estimated 90\% of the global photovoltaic market installed base \cite{Battaglia2016,Liu_2018}. Efforts to increase the efficiency of single-junction silicon solar cells are nearing the 29.43\% maximum theoretical efficiency limit \cite{Richter2013}. One proposal to surpass the theoretical efficiency limit of single-junction silicon is to create stable long-lived hot carriers by altering their thermalization and decay channels so that their excess energy may be harvested in a hot carrier solar cell \cite{Conibeer2009,Esmaielpour2018,Knig2010}. Hot carriers in semiconductors lose the majority of their heat through phonon emission \cite{Conibeer2009}, and phonon density of state engineering is considered a promising route for improved efficiency. Theoretical studies show that silicon nanoparticles exhibit a reduced phonon density of states \cite{Meier2006,Meyer2011} and lower thermal conductivity \cite{Fang2006} than bulk silicon, in particular at the lower end of the frequency spectrum \cite{Conibeer2010}. For particles less than a few hundred nm in any dimension, the low frequency acoustic phonon density of states becomes altered, preventing heat dissipation and reducing the coupling with other phonon branches \cite{Ju1999}. The loss of these low frequency acoustic phonon modes could slow the decay of the higher frequency optical phonons leading to a buildup of hot optical phonons after light excitation, a so-called ``phonon bottleneck" \cite{Conibeer2008}. Fast carrier re-excitation in silicon nanoparticles by hot phonons means that long-lived optical phonons should produce longer-lived hot carriers \cite{Prokofiev2014}. Therefore, the phonon bottleneck is hypothesized to slow both the phonon-phonon and carrier-phonon scattering.

Measuring the relationship between the phonon decay mechanisms and hot carrier lifetime in silicon nanoparticles requires simultaneous observation of electronic and structural changes over time. Extreme ultraviolet (XUV) transient absorption spectroscopy makes this possible by using a probe pulse in the XUV regime to measure core-level electronic transitions to characterize the occupancies and structural features of the valence and conduction bands. Promoting electrons from the silicon $2p$ orbital to unoccupied states creates localized core-hole excitons that impart interpretable structural information on the XUV absorption spectra \cite{Cushing2018,Seres2016}. The femtosecond resolution of the technique allows for the observation of short-lived excited carrier and phonon states \cite{Vrakking2014}.

In this study, XUV transient absorption spectroscopy is used to measure the electronic and structural dynamics in 200 nm thick single crystal silicon (100) thin films, which exhibit bulk-like properties, and 100 nm diameter, $\sim$16 nm crystalline grain size, polycrystalline silicon nanoparticles, presumed to have a confined phonon distribution. The goal is to quantify the relationship between dimensional confinement and hot carrier and phonon relaxation. Both samples are pumped with $\sim$35 fs pulses of 800 nm light, which excites electrons across the indirect transition into the conduction band $\Delta$ valley \cite{Cushing2018}, to create excited carrier densities of $2 \times 10^{19}$ cm$^{-3}$ and $6 \times 10^{19}$ cm$^{-3}$ for the nanoparticle and thin film, respectively. The measurements yield values for the carrier-phonon and phonon-phonon scattering lifetimes, which are 195 $\pm$ 20 fs and 8.1 $\pm$ 0.2 ps, respectively, for the thin film, and 870 $\pm$ 40 fs and 17.5 $\pm$ 0.3 ps, respectively, for the nanoparticles. The longer phonon-phonon lifetime for the polycrystalline nanoparticles is indicative of lower heat transport, and together with the longer carrier-phonon lifetime suggests a phonon bottleneck caused by acoustic phonon confinement. 
\section{Methods}

\subsection{Sample Preparation}

Silicon nanoparticles of nominal 100 nm diameters (undoped) with n-hexadecylamine ligands suspended in ethanol (1 mg/mL, Meliorum Technologies) were drop cast (two drops, 0.03 mg in 0.03 mL) onto 50 nm thick (3 mm $\times$ 3 mm) diamond X-ray membranes (Applied Diamond, Inc.). The thickness of the layer is estimated to be $\sim$0.05 $\mu$m based on comparison of XUV absorbance to tabulated values of silicon in literature, implying a surface coverage of approximately 50\% \cite{Henke1993}. Single crystal silicon (100) thin film samples (200 nm thick $\times$ 3 mm $\times$ 3 mm, B-doped, 10$^{15}$/cm$^{3}$, Norcada) were used as purchased; undoped silicon films were unavailable. This doping level is five orders of magnitude lower than the excitation density and should therefore have little effect on the dynamics.

\subsection{XUV Static Absorption} 

Ground state XUV spectra were measured using an XUV supercontinuum produced by high harmonic generation of near single-cycle laser pulses in neon high harmonic gas. In brief, 25 fs long carrier envelope phase stabilized pulses centered at 790 nm, produced by a 1 kHz Ti:Sapphire chirped pulse amplifier, were focused into a 1 m long hollow-core fiber filled with 1 bar of neon to broaden the spectrum by self phase modulation, covering 500 – 1000 nm, and then compressed using chirped mirrors (PC70, Ultrafast Innovations) and a 2 mm thick ammonium dihydrogen phosphate crystal to obtain $<$4 fs pulse duration \cite{Timmers2017}. The near single-cycle pulses were subsequently focused into a 1 mm long gas cell filled with 160 Torr of neon to generate the XUV supercontinuum ranging between 60 and 110 eV. The spectrometer is calibrated using M$_{4,5}$ edges of Kr \cite{Kiing1977}, L$_{2,3}$ edges of Al, and L$_{2,3}$ edges of single crystalline Si. The measurement was performed on four separate nanoparticle drop-cast samples and one thin film sample. The nanoparticle static spectrum reproduced and analyzed below is for the sample with the highest signal-to-noise ratio in the transient measurement.

\subsection{XUV Transient Absorption Spectra}

Transient XUV absorption spectra were measured using a near-infrared pump and a structured XUV continuum probe produced via high harmonic generation (HHG) with 35 fs pulses centered at 800 nm with a 32 nm bandwidth from a 1 kHz Ti:Sapphire chirped pulse amplifier \cite{Porter2018}. The setup used for transient measurements was different than that used for the static absorption due to its narrower excitation pulse bandwidth, allowing for carriers to be excited into a particular valley in the band structure. The 3.5 mJ beam produced by the amplifier is split 30/70 into a pump and probe arm, respectively, and part of the probe arm flux is also converted into 400 nm light through in-line second harmonic generation using a double optical gating apparatus \cite{Kfir2016}. The XUV probe spectrum is produced by focusing the 800 nm and 400 nm into a 40 cm long semi-infinite gas cell with approximately 250 Torr ($3.3\times10^4$  Pa) helium gas, generating even and odd harmonics from 70 – 120 eV. The XUV probe is focused onto the sample with a spot size of 200 $\mu$m and then onto a variable line space grating (35 – 110 eV), creating a dispersed spectrum that is measured by an XUV charge-coupled device camera (PIXIS-400, Princeton Instruments).

The 800 $\pm$ 16 nm pump arm is obtained directly from the beam splitter and routed through a retroreflector mounted on a delay stage to achieve time delays of -1 ps to 360 ps before passing through the sample with a spot size of 500 $\mu$m. The instrument response function at these wavelengths is approximately 50 fs. The fluence of the 800 nm pump beam was set to achieve a transient change in XUV signal between 5 – 10 mOD. The carrier density was estimated to be $2 \times 10^{19}$ carriers/cm$^3$ for the nanoparticle sample and $6 \times 10^{19}$ carriers/cm$^3$ for the thin film sample using tabulated reflectance and absorption coefficient values \cite{Cushing2018,Green2008}. To minimize thermal expansion, both samples were subjected to raster scanning (100 $\mu$m steps) between camera exposures, with each 0.3 second exposure capturing approximately 300 pulses (1 kHz repetition). To minimize laser ablation of the sample, the nanoparticles were further subjected to a dry nitrogen gas, room temperature cooling stream \cite{Porter2018}. The measurement was repeated for one thin film sample and four separate nanoparticle drop-cast samples, 2 or 3 times for each sample, and the nanoparticle measurement with the highest signal-to-noise ratio is reproduced below and used for analysis. All of the nanoparticle transient measurements captured the same qualitative signals, but differences in the local thickness of the drop cast distributions created differences in the signal-to-noise that could not be improved by averaging scans together.

\section{Results}
\begin{figure}[H]
	\includegraphics[width=1\textwidth]{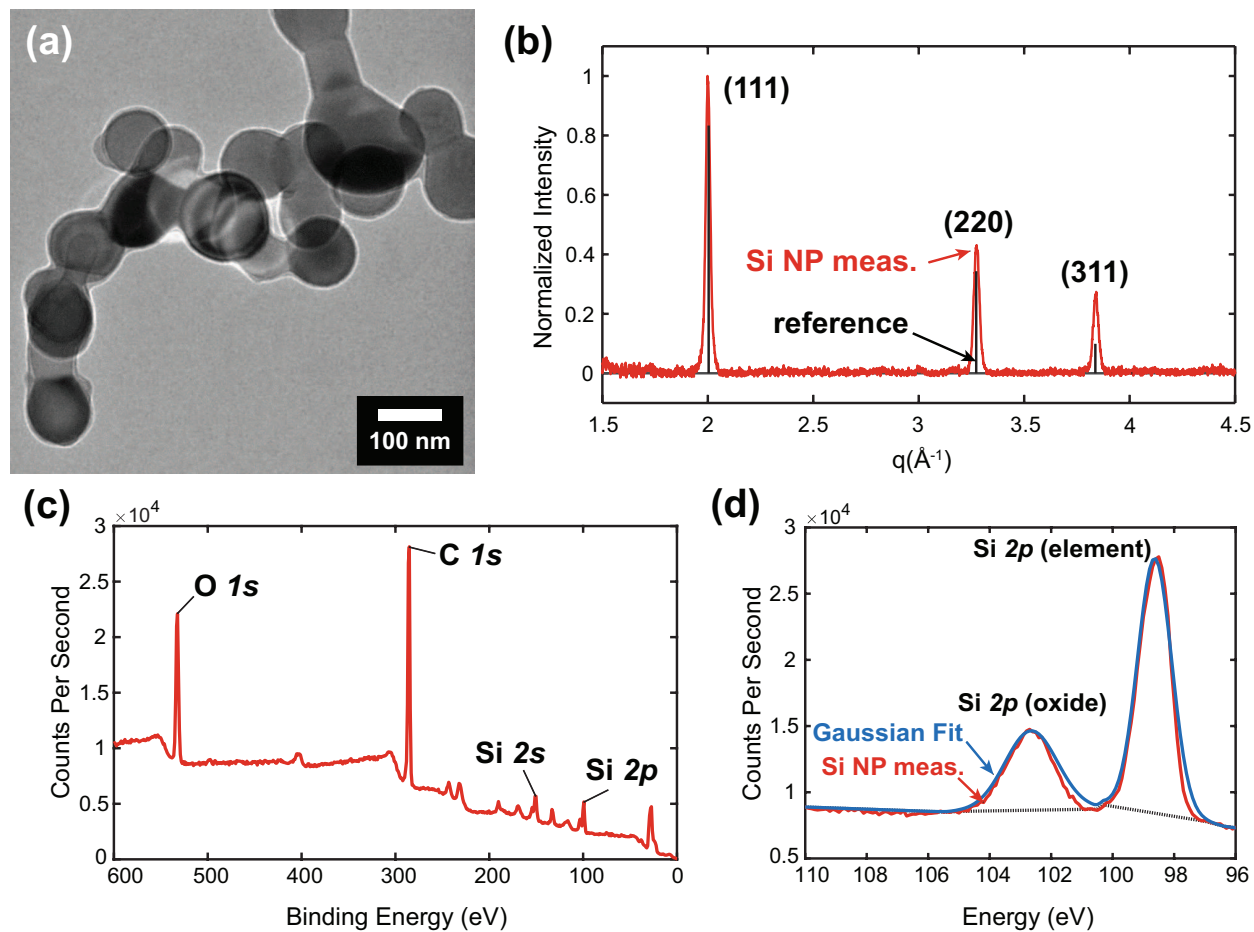}
	\caption{(a) Transmission electron microscopy image of the silicon nanoparticle sample obtained from Meliorum Technologies, which confirms the 100 nm particle diameter and non-uniform shape. (b) Powder X-ray diffraction measurement recorded here for the silicon nanoparticles (red), shown with the stick spectrum of Si taken by Downs et al. (black) \cite{Downs1993}. (c) X-ray photoelectron spectrum of the silicon nanoparticle sample. The major peaks are marked based on literature values of atomic X-ray photoelectron spectra \cite{NIST}. (d) Expanded scale elemental silicon $2p$ and silicon oxide $2p$ peaks of the X-ray photoelectron spectra (red). The peaks are fit using a Gaussian function (blue), and the background is subtracted (dotted black). }
	\label{fig1}
\end{figure}

The average diameter of the silicon nanoparticles is determined to be in the range of 100 nm from transmission electron microscopy studies (Figure 1(a)). The lack of observable lattice fringes suggests that the nanoparticles may be polycrystalline or amorphous in nature. Powder X-ray diffraction data of the nanoparticles are compared to results of previous diffraction studies performed on silicon \cite{Downs1993}, which confirms that the particles contain crystalline Si (Figure 1(b)). The average crystallite size is estimated to be 16.3 $\pm$ 1.6 nm by fitting the Si diffraction peaks at 2.001 \AA$^{-1}$, 3.274 \AA$^{-1}$, and 3.84 \AA$^{-1}$ to Gaussian functions to determine the full width at half maximum (FWHM) and then solving the Scherrer equation modified for q-space \cite{Smilgies2009}. X-ray photoelectron spectroscopy was performed (Figure 1(c)), and the Si $2p$ peak shows contributions from elemental silicon and silicon oxide (Figure 1(d)). In order to determine the thickness of the silicon dioxide layer, which is estimated to be sub-2 nm, the areas of the peaks were compared by fitting them to Gaussian functions and subtracting the background. The TEM and X-ray photoelectron measurements were performed on four nanoparticle samples, and the X-ray diffraction measurements were repeated on two nanoparticle samples, all with identical results. Similar transmission electron microscopy and powder X-ray diffraction were each performed once on the silicon (100) thin film samples (Figure S1). Electron diffraction and dark field imaging confirms that the films are single crystals with (100) orientation. Both the particles and thin film were exposed to air, so they are likely oxidized at the surface, and the major defect at the nanoparticle grain boundaries is dangling Si bonds \cite{Jackson1983}.

\begin{figure}[H]
	\includegraphics[width=0.8\textwidth]{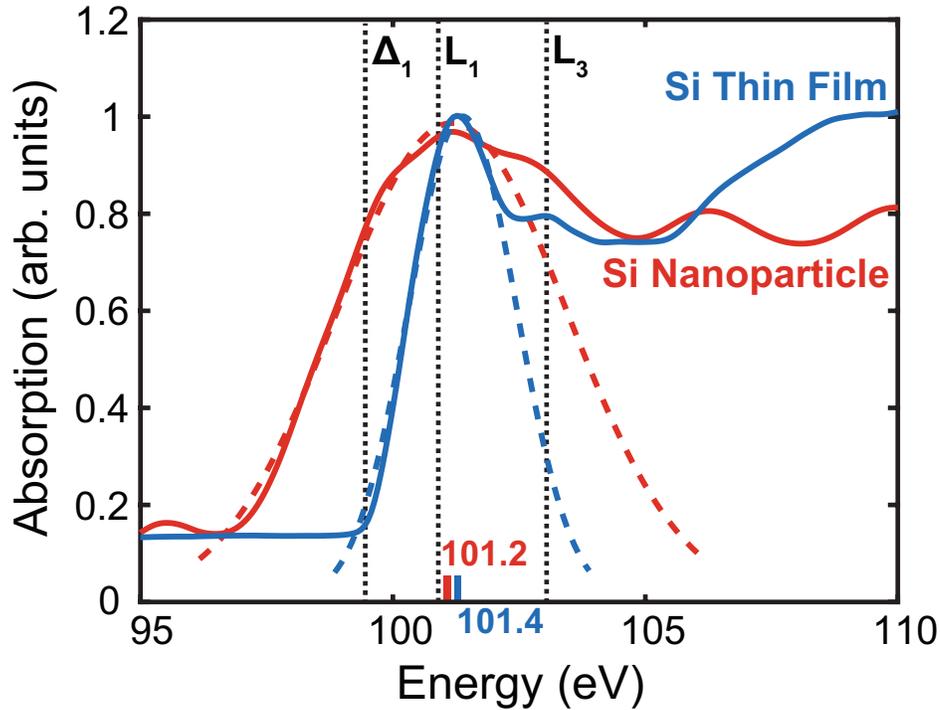}
	\caption{XUV ground state absorption spectrum of the silicon nanoparticles (red) and of the 200 nm silicon (100) thin film (blue). The initial rise in the silicon nanoparticle sample absorption versus XUV energy can be fit with a Gaussian function centered at 101.2 eV (red mark), while the initial rise in the thin film is centered at 101.4 eV (blue mark). The location of valleys in the bulk silicon band structure ($\Delta_1$, $L_1$, $L_3$) are also included (black) \cite{Cushing2018}. The absorbance intensities are normalized for visualization.}
	\label{fig2}
\end{figure}

\subsection{Static XUV Spectra}
The static XUV absorption spectrum of the silicon L$_{2,3}$ edge is measured for the silicon nanoparticles deposited on diamond X-ray membranes (red) and for the freestanding 200 nm thick (3.0 mm $\times$ 3.0 mm) single crystal silicon thin film (blue) (Figure 2). Here we discuss the differences in the static spectra between the nanoparticles and single crystalline thin film to provide information needed to interpret the transient spectra. The edge onset can be fitted with a Gaussian to the low energy rising edge of the lowest energy silicon feature (dashed red for the nanoparticles and dashed blue for the thin film). For the nanoparticles, the silicon edge occurs at $101.2 \pm 0.1$ eV (denoted by the red tick mark) and has a peak width of $\sim$5 eV, while the silicon edge of the thin film occurs at $101.4 \pm 0.1$ eV (blue tick mark) and has a width of $\sim$2.5 eV. These edge onset energies are shifted with respect to the XPS silicon peak because the transition final state is different for an absorption vs. emission process \cite{Nilsson2002}. The broad linewidths observed in both samples are the result of the extremely short core-hole lifetime of the Si $L_{2,3}$ edge, which induces broadening on the order of several eV, and by limits set by instrument resolution, on the order of tenths of an eV \cite{Cushing2018}.

Differences in properties between the two samples can explain the increased broadening and small redshift at the edge onset and the differences at higher energies observed in the XUV spectrum of the polycrystalline silicon nanoparticles. Polycrystalline Si/SiO$_{\text{x}}$ particles experience quantum confinement of carriers only at much smaller grain sizes (sub-5 nm), so a size dependent blueshift in the XUV spectrum is not expected or observed from this effect \cite{Dinh1996,Matsumoto2001}. Instead, the small domain sizes are likely responsible for the increased broadening observed in the nanoparticles due to the increased abundance of grain boundary states. For all bonding interactions involving silicon atoms at a grain boundary, the energy of the interaction depends on the geometric and orbital configurations of the atoms, which is inhomogeneous throughout the nanoparticle \cite{Jackson1983}. These variations in bond lengths and angles lead to vacant states of various energies localized at the boundary, which has been shown to increase the broadening of the $2p$ core level spectrum by nearly double the intrinsic broadening caused by state lifetime and instrument resolution \cite{Ley1982}. Furthermore, the grain boundary states, as well as the Si/SiO$_{\text{x}}$ surface states that have energies in the band gap, contribute the observed redshift \cite{Jackson1983,Sakurai1981,Kubota1999}. In related optical measurements, the absorption energy can be redshifted by tenths of an eV as the crystalline domain sizes decrease \cite{Janai1979} and defect density increases \cite{Demichellis1986}.

At higher energies above the edge onset, the silicon L$_{2,3}$ absorption spectra include contributions from the core-hole modified valence and conduction bands, as well as lattice strain and increased dielectric screening at higher energies \cite{Cushing2018}. The band structure is sensitive to bonding geometry and strain, and as such silicon atoms on the grain boundaries or surfaces of the nanoparticles will exhibit altered valence and conduction band densities of states. These in turn interact differently with the Si $2p$ core-hole, introducing features in the XUV spectrum, which may explain the appearance of the peak at $\sim$107 eV in the nanoparticle spectrum. Other explanations are surface scattering of the excited electron, or certain defect and surface states, including silicon hydride bonds \cite{Ley1982} and silicon oxide bonds in the sub-2 nm silicon oxide layer, which may open up states above the conduction band.

\begin{figure}[H]
	\includegraphics[width=1\textwidth]{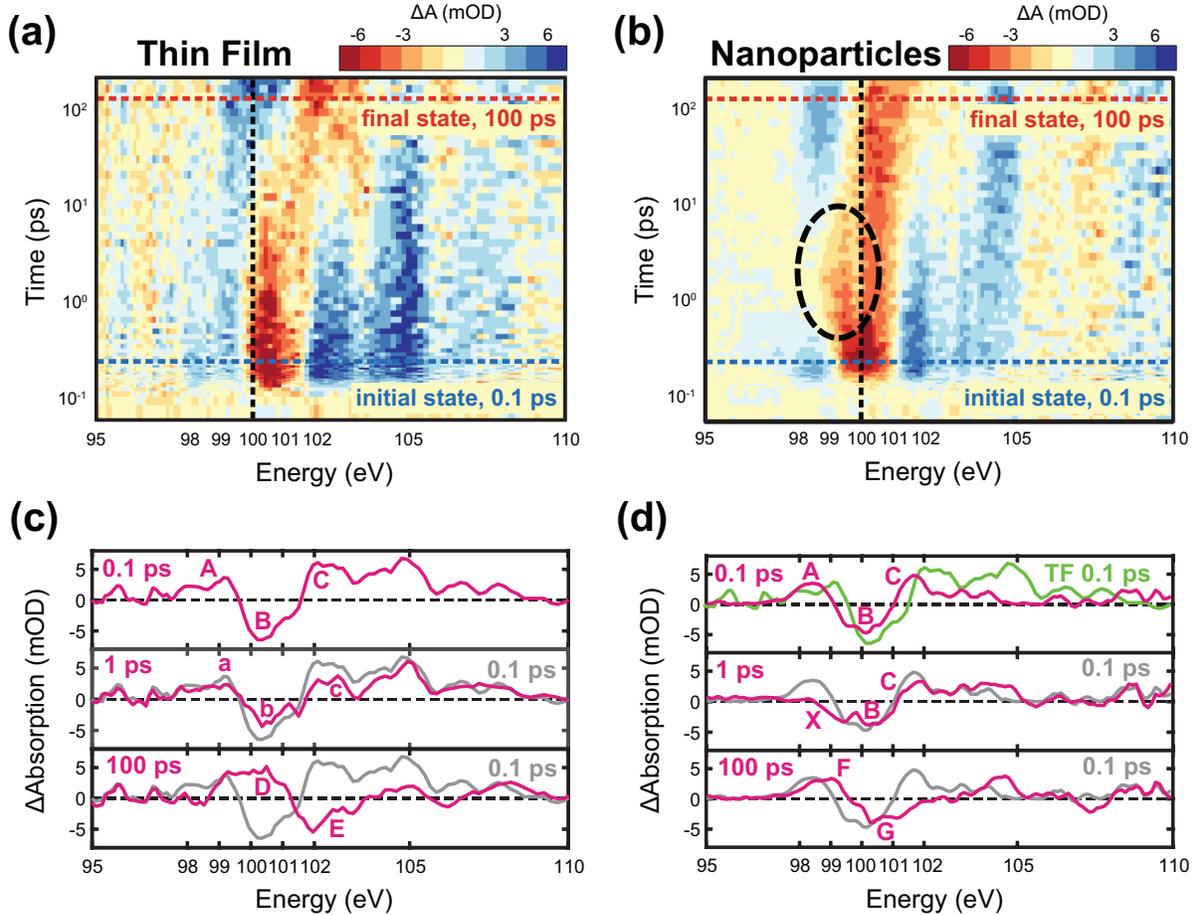}
	\caption{Transient XUV differential absorption spectra of the (a) silicon thin film and (b) silicon nanoparticles after 800 nm photoexcitation for the first 200 ps. The transient maps are shown with a logarithmic time axis, offset by +100 fs for visual clarity. The horizontal dotted lines represent the states used for multivariate regression analysis, with the dotted blue line representing the ``initial state" at 0.1 ps and the dotted red representing the ``final state" at 100 ps. The vertical dotted lines at 100 eV help visualize small shifts in the features. The feature in (b) highlighted by the dashed black circle is explained in the text. Lineouts at three important timescales are presented below each figure to highlight the spectral features of the (c) thin film and (d) nanoparticles, which are labeled in magenta letters and detailed in the text. Each of the magenta lineouts, taken at approximately 0.1 ps, 1 ps and 100 ps after excitation, is the average of the nearest five time slices. In the middle and bottom panels, the 0.1 ps lineout for each respective sample is displayed in gray to visualize changes to the differential spectra. In the top panel of (d), the green line labeled `TF 0.1 ps' is the thin film differential spectrum at 0.1 ps.}
	\label{fig3}
\end{figure}

\subsection{Transient XUV Spectra}
The XUV transient differential absorption spectra after 800 nm optical excitation were measured for the silicon thin film (Figure 3(a)) and silicon nanoparticles (Figure 3(b)) from 0 to 200 ps. An 800 nm excitation is chosen to match the energy of the indirect transition into the $\Delta$ valley, which is the lowest energy band structure critical point, because this keeps the relaxation dynamics simple by eliminating inter-valley relaxation \cite{Cushing2018}. Carriers are excited at the $X$ symmetry point with an excess energy of 0.35 $\pm$ 0.03 eV above the $\Delta_1$ conduction band minimum. The samples are excited to similar carrier densities, approximately $2 \times 10^{19}$ cm$^{-3}$ for the nanoparticles and $6 \times 10^{19}$ cm$^{-3}$ for the film, to ensure that similar dynamics are being compared. Higher carrier densities are not used because the nanoparticles cannot dissipate the excess heat and destroy the substrate membrane at densities above $1 \times 10^{20}$ cm$^{-3}$. The transient colormaps are plotted here with a logarithmic time axis to show more clearly the changing spectral features at early times, but the logarithmic time axis obscures timescales and makes it difficult to visually compare the decay rates. To highlight certain features, spectral lineouts are plotted (Figure 3(c),(d)) of the average of the five time points surrounding 0.1 ps, 1 ps, and 100 ps. The most prominent effects on the transient spectra occur in the near-edge region (98 – 102 eV).

At 0.1 ps, immediately after photoexcitation, the thin film differential spectrum (Figure 3(c), top panel) has an increase in absorption centered at 99 eV labeled feature A, a decrease in absorption at $\sim$100.5 eV labeled feature B, and positive signals above 102 eV labeled feature C. In the nanoparticle spectrum taken at 0.1 ps after excitation (Figure 3(d), top panel, magenta), the same three features are also observed, albeit with a slight redshift and less intensity in the $>$100 eV features. To compare these two spectra more easily, the thin film 0.1 ps spectrum is plotted in the top panel of Figure 3(d) in green, labeled `TF 0.1 ps'. The increase in absorption labeled A that occurs at 99 eV in the thin film is closer to 98.5 eV in the nanoparticles, and the negative peak labeled B is located at 100.5 eV in the thin film and $\sim$100 eV in the nanoparticles. This redshift between the nanoparticle and thin film differential spectra is expected due to the initially red-shifted and broadened onset of the absorption peak in the static ground state spectrum of the nanoparticles (Figure 2). Additionally, the differences in the higher energy features between the thin film and nanoparticles may be explained by the differences in the static spectra.

At 1 ps, an overall decrease in the magnitude of the features is observed in the thin film transient spectrum (Figure 3(c), middle panel, magenta) when compared to the spectrum at 0.1 ps (gray). The spectral features are therefore labeled with lowercase letters a, b and c to denote the intensity decrease. On the other hand, the nanoparticle transient spectrum at this time (Figure 3(d), middle panel, magenta) exhibits a broadening on the low energy side of the negative feature centered at $\sim$100 eV, labeled X, when compared to the 0.1 ps spectrum (gray), and there is no decrease in the magnitude of the higher energy B and C features. This new low energy feature, indicated on the transient colormap (Figure 3(b)) by the black dashed circle, appears as a new decrease in absorption at $\sim$99 eV, and then it disappears from the spectrum within 10 ps. By 100 ps, the thin film transient signal (Figure 3(c), bottom panel, magenta) exhibits a positive feature at $\sim$100.5 eV labeled feature D, and a negative feature at $\sim$103 eV labeled feature E. These features appear at the same energies as the B and C features in the 0.1 ps spectrum (gray), respectively, but have opposite signs. In the nanoparticle transient signal at 100 ps (Figure 3(d), bottom panel, magenta), large changes in the intensity and sign of the peaks are not observed when compared to the 0.1 ps spectrum (gray). The near-edge features from 98 eV – 102 eV, labeled F for the positive signal at 99 eV and G for the negative signal at 100.5 eV, exhibit a $\sim$0.5 eV blueshift from the A and B features of the 0.1 ps spectrum. These shifts are significantly smaller than the $>$2 eV shift of the negative peak position in the thin film spectra from the B feature at 0.1 ps to the E feature at 100 ps (Figure 3(c), bottom panel).

Before assigning these observed spectral changes to processes in the samples, we first present the expected dynamics for a crystalline silicon sample. After photoexcitation into the $\Delta$ valley, charge carriers in single crystal silicon quickly form a hot thermal population within the 35 fs excitation pulse duration \cite{Schultze2014}. The new charge distribution in the conduction and valence bands blocks certain transitions and opens others, known as state filling, and the resulting Coulombic forces between the charged bands alters their energies, causing broadening and band gap renormalization \cite{Cushing2018}. Within the first few-hundred femtoseconds, hot electrons decay to the $\Delta$ valley edge via scattering with optical phonons \cite{Sabbah2002}, creating a bath of hot optical phonons with a 100 – 250 fs scattering timescale \cite{Lee2005,Letcher2007}. The time resolution of the experiment prevents observation of the hot carrier population before it begins to scatter with optical phonons, so the initially measured spectrum has both a hot carrier and a hot optical phonon population. After a few picoseconds, the charge carrier population is fully thermalized to the band edge, leading to diminished effects of state filling, broadening, and renormalization, while phonon-phonon scattering of energy from the optical phonon branches to the lower-frequency acoustic phonon branches has begun, with a 2 – 10 ps acoustic phonon scattering time \cite{Lee2005,Harb2006}. By 100 ps, almost all the excess energy pumped into the system has been funneled into the hot acoustic phonon population, which corresponds to heating of the lattice, and this state persists for nanoseconds \cite{Cuffe2013}. In summary, at early timescales ($\sim$ 0.1 ps), the primary contributions to the transient response are electronic and hot optical phonon effects, and after about 10 ps the response is mainly caused by hot acoustic phonons and the heated lattice.

The XUV transient absorption spectrum of the silicon thin film at the L$_{2,3}$ edge can be described by three spectral components caused by the excited carriers, hot optical phonons and hot acoustic phonons, as demonstrated in previous studies \cite{Cushing2018,Cushing2019,Cushing2020} and reproduced in Figure S2 for a carrier density of $1 \times 10^{19}$ cm$^{-3}$. The hot optical phonon effects are replicated when the DFT-based model introduced in Cushing et al.\cite{Cushing2018} includes an anisotropic lattice expansion in the [100] direction, which is the expected symmetry of intra-valley scattering between the degenerate $\Delta$ valleys. Similarly, the effects on the spectrum caused by the hot acoustic phonons can be modeled with a symmetric isotropic lattice expansion ([111]), which would occur due to heating. The near-edge region contributions from both the electronic and hot optical phonon effects increase the absorption (positive signal) below 100 eV and decrease the absorption (negative signal) between 100 eV and 102 eV (Figure S2(a)), which is the same spectral shape as is observed in the 0.1 ps spectra of both samples (Figure 3(c) and 3(d), top panels). Therefore, both samples exhibit hot carrier and hot optical phonon populations at 0.1 ps after photoexcitation. The acoustic phonon and heated lattice contributions to the transient spectra instead increase absorption (positive signal) below about 101 eV and decrease absorption (negative signal) between 101 eV and 106 eV (Figure S2(b)), which is the same spectral shape as the 100 ps thin film spectrum (Figure 3(c), bottom panel). Thus, the thin film sample used here exhibits a hot acoustic phonon population 100 ps after photoexcitation. Additionally, the 1 ps spectrum of the thin film has the same spectral shape as the hot carriers and hot optical phonons, but with decreased intensity. At this timescale, the photoexcited carriers are expected to have mostly decayed via phonon emission \cite{Prokofiev2014,Lee2005}, so a decrease in the intensity of spectral features is expected, as the increasing contributions from acoustic phonons cancel out the decreasing contributions from electronic changes and optical phonons \cite{Cushing2018}.
 
The assignment is more complex for the nanoparticle sample after 1 ps. The intensities of the features associated with hot carriers and hot optical phonons do not decrease by this timescale, implying a slower transfer of energy from the hot carriers and hot optical phonons to the acoustic phonons, and a new negative feature centered at $\sim$99 eV appears. Because this feature occurs at energies corresponding to within the band gap and is distinct from the signals caused by the core-hole mediated band gap renormalization and broadening observed at 0.1 ps, it is attributed to electrons filling gap states. As mentioned above, these states are likely to be at the grain boundaries and Si/SiO$_{\text{x}}$ interfaces and surfaces. A more complete assessment of this feature is included in the Discussion. By 100 ps, the nanoparticle transient spectrum does not match the modeled spectral components of hot acoustic phonons from Cushing et al. (Figure S2(b))\cite{Cushing2018}, and instead the features appear midway between the acoustic phonon features and the 0.1 ps nanoparticle spectrum containing the electronic and optical phonon features. The negative feature that blueshifts by $>$2 eV between 0.1 ps (feature B at 100.5 eV) and 100 ps (feature E at 103 eV) in the thin film (Figure 3(c), bottom panel) only blueshifts by $\sim$0.5 eV over the same time interval in the nanoparticle, from feature B at 100 eV to feature G at 100.5 eV (Figure 3(d), bottom panel). This implies that the 100 ps nanoparticle spectrum is caused by a combination of the three spectral components identified previously for excited carrier, hot optical phonons and acoustic phonons \cite{Cushing2018}. Thus, the nanoparticle spectrum at 100 ps can be explained as the sum of the signals caused by hot acoustic phonons plus hot carriers, hot optical phonons, or both. This indicates that, for the nanoparticle sample, long-lived hot carriers or hot optical phonons may be present for up to 100 ps because their contributions to the signal are still prominent at these timescales.

\begin{figure}[H]
	\includegraphics[width=1\textwidth]{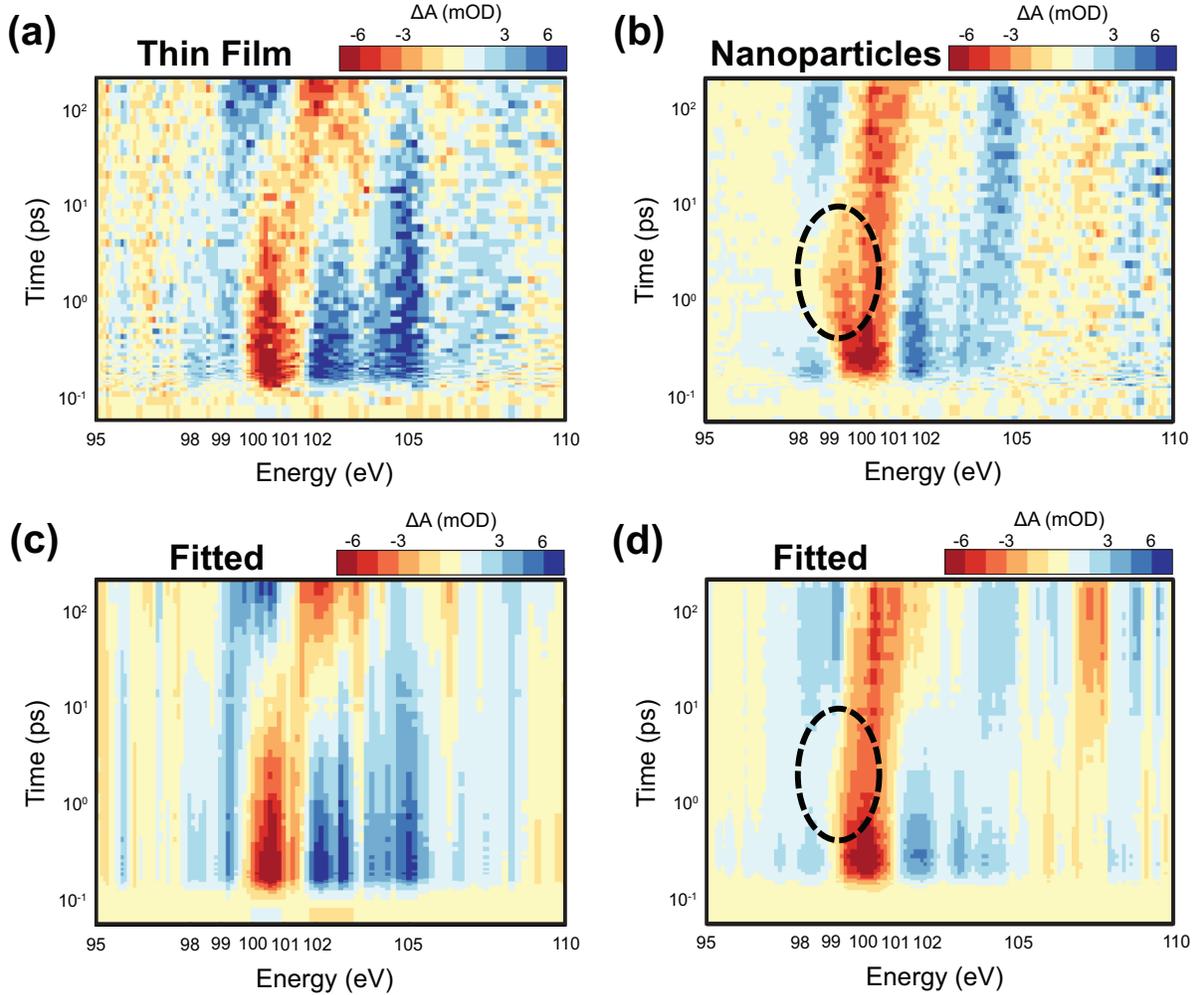}
	\caption{(a) and (b) show the raw transient XUV differential absorption spectra for the thin film and nanoparticle samples, respectively. These colormaps are identical to those shown in Figure 3(a) and (b) but without the guidelines. (c) and (d) show the results, for the thin film and nanoparticle, respectively, of the multivariate decomposition of the transient spectra into contributions from the initial and final states. Most of the features in the raw data are reproduced, except the low energy broadening of the negative feature in the nanoparticle sample, indicated by the black dashed circles. The transient maps are shown with a logarithmic time axis out to 200 ps, offset by 100 fs for visual clarity.}
	\label{figs3}
\end{figure}
\subsection{Kinetic Modeling of the Dynamics}
To quantify the timescales of the ultrafast carrier and phonon relaxation dynamics observed in the single crystal and nanoparticle samples, a multivariate regression is performed to decompose the complex transient spectra into contributions from an initial state, with hot carriers and hot optical phonons, and a final state, with a hot acoustic phonon population. These states are chosen at 0.1 ps and 100 ps timescales, respectively, which are indicated by the blue (initial) and red (final) dotted lines in Figure 3. The results of the regression fit are shown as colormaps in Figure 4 for the (c) thin film and (d) nanoparticles, plotted along with the raw data in panels (a) and (b), respectively.  The amplitudes obtained for these states over time are denoted by the box points (blue for initial state, red for final state) in Figure 5(a) for the thin film and 5(b) for the nanoparticles. This decomposition into two states, instead of three for the three hot populations (carriers, optical phonons, and acoustic phonons), is justified from the similarity between the spectra associated with hot carriers and hot optical phonons \cite{Cushing2018}, as well as the difficulty in separating the states at early times due to the limited time resolution. Although the nanoparticle state at 100 ps contains signals of all three hot populations, it persists and is unchanging for the entire 300 ps delay stage, so it is sufficient here to use this as the final state in the fit. The decrease in absorption observed near 99 eV at about 1 ps in the nanoparticle sample (black dashed circle in Fig. 3(b)) is likely the result of grain boundary defect and surface carrier relaxation pathways in the polycrystalline nanoparticles. As will be discussed below, the density of these defect states is an order of magnitude below the carrier excitation density, which is too low to affect the overall carrier thermalization. Therefore, carrier decay via trap states is not included in this analysis, since neither the initial 0.1 ps spectrum nor the final 100 ps spectrum contain this negative feature, as can be seen by comparing the raw transient colormap with the multivariate regression fit (Figure 4).

The obtained regression amplitudes are then fit to a model to describe the carrier excitation, optical phonon scattering, and acoustic phonon scattering dynamics in both the thin film and nanoparticle samples. For laser heated silicon, carrier and lattice decay dynamics are typically fit using a three-temperature kinetic model, in which the speed of the exchange between the three excited populations is based on the relative temperatures of the populations and the lifetime of the scattering modes \cite{Cushing2018,Lee2005}. The heat exchange between two populations is therefore dependent on the difference in temperature between these populations, $(T_1 (t)-T_2 (t))$, the heat capacity, $C_1$, and the scattering time, $\tau_{12}$. Thus, the three temperature model is given by: 

\begin{equation}
	 C_e \frac{\partial T_e(t)}{\partial t} = N(exc) - \frac{C_e(T_e(t) - T_o(t))}{\tau_{eo}}\label{eqn1}
\end{equation}
\begin{equation}
	C_o \frac{\partial T_o(t)}{\partial t} = \frac{C_e(T_e(t) - T_o(t))}{\tau_{eo}} - \frac{C_o(T_o(t)-T_a(t))}{\tau_{oa}}\label{eqn2}
\end{equation}
\begin{equation}
	C_a \frac{\partial T_a(t)}{\partial t} = \frac{C_o(T_o(t)-T_a(t))}{\tau_{oa}}\label{eqn3}
\end{equation}
Where $t$ is time, $T$ is the population temperature and $C$ is the heat capacity, with the subscript $e$ for hot carriers, $o$ for the optical phonons, and $a$ for the acoustic phonons. Values for the heat capacities are taken from Lee\cite{Lee2005}. The first term in Equation (1), $N(exc)$, accounts for the initial excitation of hot carriers by the laser pulse envelope \cite{Cushing2018}. The two fitted values in this model are $\tau_{eo}$, the lifetime of energy loss from hot carriers to optical phonons, and $\tau_{oa}$, the lifetime of energy loss from optical phonons to acoustic phonons \cite{Lee2005}. Direct decay of hot carriers via acoustic phonon emission is not included because the addition of this term has no effect on the fitted carrier-optical and optical-acoustic decay times. Additional processes, which are slow or have a low likelihood of occurring and are therefore below the noise floor of this measurement, have been left out of this model to allow for fewer fit parameters to be used. Auger recombination of conduction band electrons with valence holes is not included because the Auger timescale at this carrier density is very long ($>$100 ps )\cite{Richter2012}. Impact ionization, in which hot carrier scattering excites new hot charge carriers, is omitted because it has been shown to have an insignificant effect on the Si L$_{2,3}$ edge for similar pump fluences \cite{Cushing2019}. Thermal diffusion of carriers and lattice heat is ignored, as the addition of these terms to the model had no effect on the calculated decay times. 

To apply the three-temperature model to the XUV transient spectral data, a few further modifications are included. Since there are only two states used in the multivariate regression analyses, and since the initial states at 100 fs likely comprise signals from both hot carriers and optical phonons, the amplitude of the initial state over time is treated as the linear combination of both states in the three-temperature model fit. Additionally, the average population temperatures are unknown and only the amplitudes of the regression states, $\eta$, are known. Thus, the regression amplitudes for each population, multiplied by a population weighting factor, are used instead of the temperature, as has been applied in Carneiro et al \cite{Carneiro2017}. This is an acceptable simplification when solving for the average timescale of energy transfer between energy- and momentum-averaged populations, and not an energy- or momentum-specific timescale. The three-temperature model, modified to use the amplitude of the three populations instead of the temperature, is as follows:

\begin{equation}
	C_e \frac{\partial \eta_e(t)}{\partial t} = N(exc) - \frac{C_e(\eta_e(t) - \eta_o(t))}{\tau_{eo}}\label{eqn4}
\end{equation}
\begin{equation}
	C_o \frac{\partial \eta_o(t)}{\partial t} = \frac{C_e(\eta_e(t) - \eta_o(t))}{\tau_{eo}} - \frac{C_o(\eta_o(t)-\eta_a(t))}{\tau_{oa}}\label{eqn5}
\end{equation}
\begin{equation}
	C_a \frac{\partial \eta_a(t)}{\partial t} = \frac{C_o(\eta_o(t)-\eta_a(t))}{\tau_{oa}}\label{eqn6}
\end{equation}
Where $\eta_e (t)$ is the hot carrier state amplitude, $\eta_o (t)$ is the optical phonon state amplitude, $\eta_a (t)$ is the acoustic phonon state amplitude. The total initial state fitted amplitudes, comprising the sum of the hot carrier and optical phonon states, are given by the blue lines in Figure 5(a) and 5(b), while the final state amplitude fits to the acoustic phonon state are the red lines. The acoustic phonon state amplitude appears to reach a plateau within 10 ps in the thin film sample but continues to increase beyond 20 ps in the nanoparticles. As seen in Figure 5(c) for the thin film and 5(d) for the silicon nanoparticle sample, the total fitted initial state (solid) is decomposed into the contributions from hot carriers (dashed) and optical phonons (dotted). The hot carrier state dominates early timescales of less than 1 ps before decaying, while the optical phonon appears and reaches a maximum within 1 ps for the thin film, but only past 2 ps for the nanoparticles, before also decaying.

\begin{figure}[H]
	\includegraphics[width=1\textwidth]{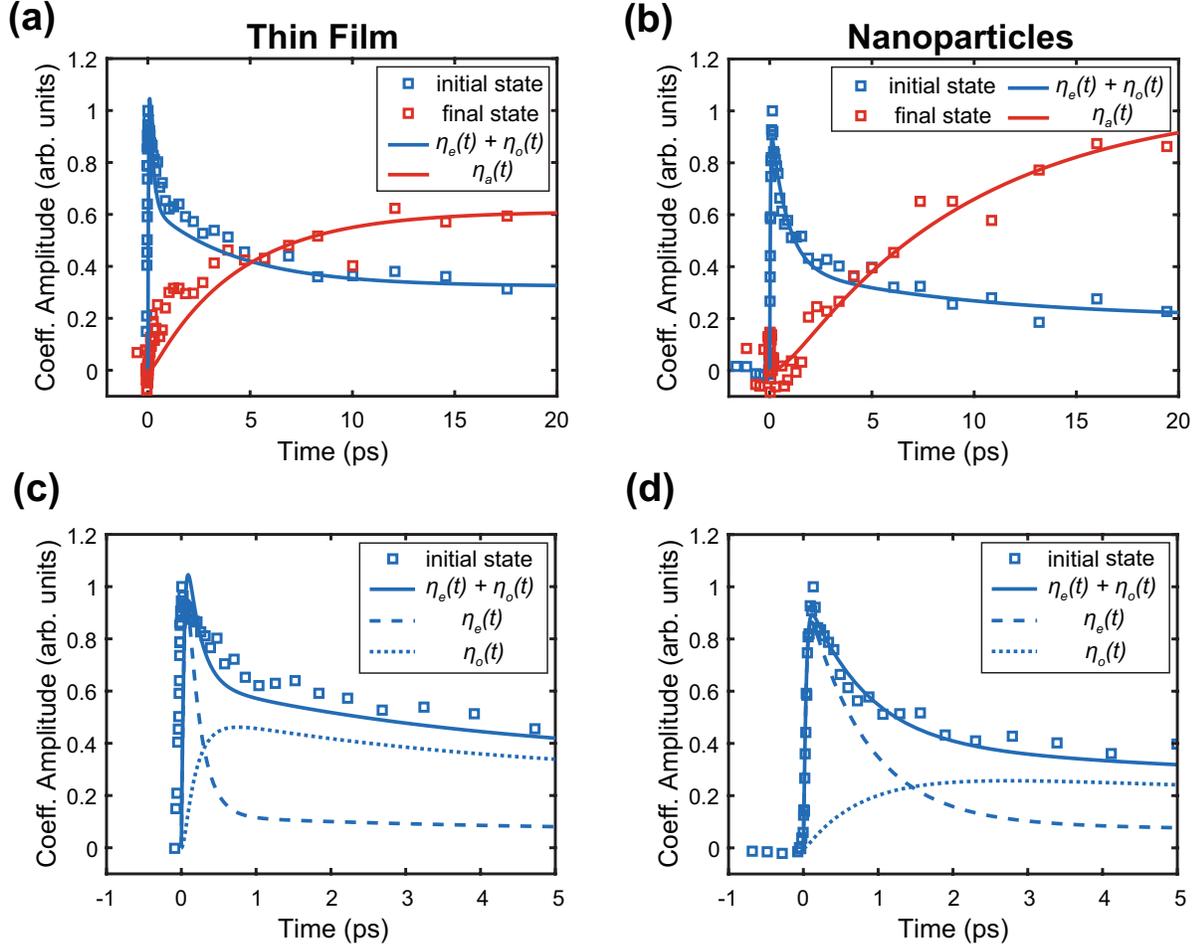}
	\caption{Amplitudes of the multivariate regression performed on both (a) the silicon thin film and (b) the nanoparticle sample for the first 20 ps are shown as the box points, blue for the initial state at 0.1 ps and red for the final state at 100 ps. The fit using the kinetic model is indicated by solid lines. The blue lines indicate the sum of the fitted amplitudes of the hot carrier state and the hot optical phonon state, $\eta_e (t)+\eta_o (t)$, and the red lines correspond to the fitted amplitude of the acoustic phonon state, $\eta_a (t)$. Close-up early time of the initial state regression amplitudes and fits of the silicon thin film (c) and nanoparticles (d) are plotted for the first 5 ps to improve clarity. Boxes indicate the multivariate regression amplitude of the initial state, and the solid line is the sum $\eta_e (t)+\eta_o (t)$. The fit is decomposed into contributions from the hot carrier state fit $\eta_e (t)$ (dashed) and optical phonon state fit $\eta_o (t)$ (dotted).}
	\label{fig4}
\end{figure}

Using the above model, $\tau_{eo}$ was extracted to be 195 $\pm$ 20 fs and 870 $\pm$ 40 fs for the thin film and nanoparticle samples, respectively, via a global fitting with multiple starting guesses (MATLAB 2018b, MultiStart). The error indicated is one standard error of the fit. $\tau_{oa}$ was calculated to be 8.1 $\pm$ 0.2 ps and 17.5 $\pm$ 0.3 ps for the thin film and nanoparticle samples, respectively. The values calculated for the thin film sample agree with previously reported values of approximately 100 – 250 fs for $\tau_{eo}$  \cite{Lee2005,Letcher2007,Sjodin1998} and approximately 2 – 10 ps for $\tau_{oa}$ \cite{Lee2005,Harb2006}. Even considering the range of values in various measurements, and the large uncertainty inherent in the many assumptions of this model, the lifetimes of optical phonon scattering and acoustic phonon scattering are much longer in the silicon nanoparticles. Furthermore, the slightly larger carrier excitation density used for the thin film sample ($6 \times 10^{19}$ cm$^{-3}$ vs. $2 \times 10^{19}$ cm$^{-3}$ for the nanoparticles) should have caused a longer thin film scattering time [45]. This indicates that both hot carriers and hot optical phonons are much longer-lived in dimensionally confined polycrystalline silicon compared to single-crystalline bulk.

\section{Discussion}
Now we consider the reasons for the slowed carrier and optical phonon cooling in the nanoparticle sample. One major difference between the two samples is crystallinity, but as explained in the following calculation, the higher abundance of defects and grain boundary surface states in the nanoparticles is still much smaller than the carrier excitation density, and therefore this is not a plausible explanation for the longer-lived hot carriers and hot optical phonons in the nanoparticles. Following the model of Ref. 46 \cite{Amit2014}, the known grain boundary surface trap state density of silicon \cite{Seto1975,DeGraaff1982}, and the XRD-measured nanoparticle grain sizes of 16.3 $\pm$ 1.6 nm, the nanoparticle sample used here has a calculated grain boundary trap state density of $1$ $–$ $4 \times 10^{18}$ cm$^{-3}$. The excited carrier density used is approximately $2 \times 10^{19}$ carriers/cm$^3$, indicating that these trap states are not a majority decay channel. Instead, carrier trapping at defects and subsequent recombination occurs concurrently with phonon emission and is responsible for the distinct signal at $\sim$99 eV and 1 ps in the nanoparticle transient spectrum (dashed black circle in Figure 3(b), X feature in Figure 3(d)). As discussed above, the location and sign of this feature is consistent with electrons filling vacant states right below the conduction band edge, which is the energy of the grain boundary surface states \cite{Cushing2018}. Carrier trapping into these states, which have energies in the band gap due to vacancies and the inhomogeneous bonding geometries at grain boundaries, occurs within $\sim$0.6 – 2.0 ps in nanocrystalline silicon \cite{Esser1990,Fekete2009}. Similarly, amorphous silicon exhibits a 0.8 – 1 ps carrier thermalization time and a 10 – 30 ps recombination lifetime \cite{Fauchet1989,Titova2016}. This $\sim$1 ps trapping timescale and $\sim$10 ps recombination is similar to the timescales of the appearance and disappearance of the negative spectral feature at $\sim$99 eV (Figure 3(b), black dotted circle). Exponential fitting of the growth and decay of the feature, corresponding to the amplitude of the 1 ps time delay (Figure 3(d), middle panel), yields a growth time of 570 $\pm$ 60 fs and a decay time of 11.2 $\pm$ 3.4 ps (Figure S3). While the carrier trapping timescales are very similar to the 870 $\pm$ 40 fs timescale of the nanoparticles measured here, the density of midgap grain boundary surface states is too low to attribute this timescale to trapping. Furthermore, trapping in the midgap states is unlikely to alter the measured carrier-phonon scattering in the bulk. Transient reflectivity studies of polycrystalline silicon films of different nano-sized domains ranging from 9 – 19 nm have shown that polycrystallinity alone does not alter carrier cooling times \cite{Myers2001}. While those films exhibited a long-lived signal from the grain boundary surface states, the carrier decay rate by phonon emission was identical to crystalline silicon \cite{Myers2001}. Thus, while carrier trapping by grain boundary surface states occurs sufficiently to contribute to the observed transient XUV signals, it is not the majority channel for excited carrier thermalization, and thus it is not the cause of the slowed hot carrier and hot phonon decay.

It should be noted that the approximately $10^{18}$ cm$^{-3}$ defect density calculated for these nanoparticles is larger than the typical photoexcited carrier densities observed in a functioning silicon photovoltaic device under standard solar illumination \cite{Uprety2019}. In order for grain boundary surface defect trapping and recombination to remain a minority carrier decay channel in these devices, silicon nanoparticle samples with larger crystalline grains, and therefore fewer defect states, must be considered in future XUV transient absorption studies.

Now we consider the explanation for the longer scattering lifetimes measured in the silicon nanoparticles in terms of the creation of a ``phonon bottleneck" due to confinement of acoustic phonon modes \cite{Conibeer2010,Conibeer2008}. For 100 nm particles or less like the ones studied, the acoustic phonon density of states (PDOS) is altered, leading to reduced thermal transport and diminished coupling with other phonon branches \cite{Ju1999}. When photoexcitation of these particles produces hot optical phonons, their thermalization via acoustic phonon scattering is slowed, leading to higher energy optical phonon states being occupied for longer periods of time. This effect is not expected in the 200 nm $\times$ 3 mm $\times$ 3 mm thin film because only one dimension is small enough to have an altered PDOS, and therefore only one of the three acoustic phonon branches is reduced. In silicon, hot optical phonons can easily impart their excess energy back into carriers, a process called phonon recycling \cite{Prokofiev2014}; thus long-lived hot optical phonons translate to long-lived hot carriers. In this scenario, acoustic phonon confinement increases both the phonon-phonon and carrier-phonon scattering times by re-excitations, as the heat imparted by the laser cannot be efficiently removed via thermalization. This effect has been observed in 70 nm $\times$ 200 nm single-crystal silicon nanopillars, in which the electron-phonon scattering time after 800 nm excitation is increased to 400 fs \cite{Chekulaev2018}. The even longer 870 $\pm$ 40 fs carrier lifetime measured in this study is expected as the 100 nm nanoparticles and 16 nm grains are smaller. Moreover, the longer phonon-phonon scattering lifetime measured here indicates decreased heat transport, consistent with a diminished acoustic phonon bath. Thus, the phonon bottleneck hypothesis is consistent with all of the observed dynamics.

An additional effect that may be occurring in the nanoparticle sample is confinement of the optical phonons, which occurs for particles and grains of less than 25 nm \cite{Meier2006}. Although the silicon nanoparticles studied here have nominal 100 nm diameters, the PDOS approaches zero at grain boundaries, and therefore phonons may become confined in the 16 nm grains \cite{Camus2008}. The result of a confined optical phonon bath is slowed carrier thermalization, which is the same result observed when a hot phonon bottleneck occurs. Thus, a reduced optical phonon density of states may be present in the nanoparticles, but its effect on the carrier thermalization cannot be confirmed in this study.

\section{Conclusions}
The carrier dynamics of a single crystal silicon thin film and of polycrystalline silicon nanoparticles were measured using XUV transient absorption spectroscopy. The carrier-phonon scattering and phonon-phonon scattering lifetimes for both the thin film and the nanoparticles were calculated using a three-temperature kinetic model, with coefficients obtained by a multivariate regression of the transient spectra. The carrier-phonon and phonon-phonon scattering lifetimes for the silicon nanoparticles (870 $\pm$ 40 fs and 17.5 $\pm$ 0.3 ps, respectively) were much longer than that for the silicon thin film (195 $\pm$ 20 fs and 8.1 $\pm$ 0.2 ps, respectively). In agreement with the phonon bottleneck hypothesis, carrier-phonon and phonon-phonon scattering in the nanoparticles is slowed, and this slowing is unlikely to be the result of increased surface states and defects in the polycrystalline nanoparticles. These results support the hypothesis that there is a reduction of the low frequency and low energy acoustic phonon mode density of states, which severely limits the optical-to-acoustic phonon scattering, and therefore the initial carrier thermalization. Further exploration into the carrier and phonon lifetimes of different sizes and shapes of nano silicon, possible simultaneously with XUV transient absorption spectroscopy, may provide further confirmation of slowed carrier cooling when the silicon phonon bath is confined.

\begin{acknowledgement}

This work was supported by the U.S. Department of Energy, Office of Science, Office of Basic Energy Sciences, Materials Sciences and Engineering Division, under Contract No. DE-AC02-05-CH11231 within the Physical Chemistry of Inorganic Nanostructures Program (KC3103).  S.K.C. acknowledges support by the Department of Energy, Office of Energy Efficiency and Renewable Energy (EERE) Postdoctoral Research Award under the EERE Solar Energy Technologies Office. H.-T. C. acknowledges support from Air Force Office of Scientific Research (AFOSR) (FA9550-15-1-0037 and FA9550-19-1-0314) and W. M. Keck Foundation (No. 046300).  J.C.O. gratefully acknowledges the support of the Kavli Energy NanoScience Institute / Philomathia Graduate Student Fellowship. Work at the Molecular Foundry was supported by the Office of Science, Office of Basic Energy Sciences, of the U.S. Department of Energy under Contract No. DE-AC02-05CH11231. The authors gratefully acknowledge mentorship and guidance from Lucas M. Carneiro.

\end{acknowledgement}

\begin{suppinfo}
	Description of TEM, XPS, and powder XRD measurements; figures of thin film characterization, modeled differential Si $L_{2,3}$ edge, and growth of trap state feature.
\end{suppinfo}

\bibliography{sinp_jpcc_ref}

\begin{figure}[H]
	\includegraphics[]{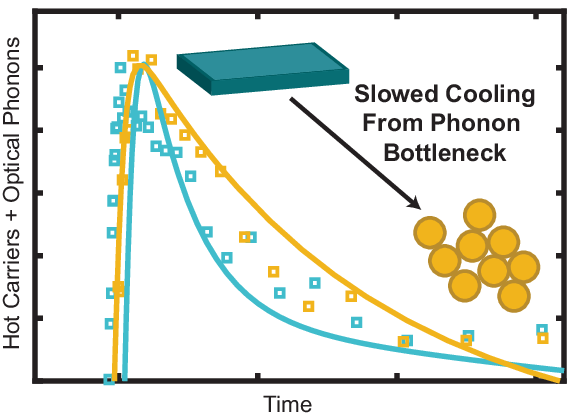}
	\caption{TOC Figure}
	\label{toc}
\end{figure}

\end{document}